\documentclass[a4paper]{elsart}
\newcommand{\brochette}{\cite{BHP05,CHM02,CH94,DGMcC01,EGMcCS94,HMPV00,HPV99,McCS99}\xspace}

\usepackage{dam06}
\newtheorem{proposition}{Proposition}
\newtheorem{definition}{Definition}
\newtheorem{lemma}{Lemma}
\newtheorem{corollary}{Corollary}
\newtheorem{theorem}{Theorem}
\newcommand{\remark}{\emph{Remark:~~}}
\newenvironment{proof}{\begin{trivlist}
 \item[\hspace{\labelsep}{\em\noindent Proof: }]
}{\hfill$\Box$\end{trivlist}}
\begin{document}
\begin{frontmatter}
\title{Algorithmic Aspects of a General Modular Decomposition Theory}
\author[lirmm]{B.-M. Bui-Xuan}
\author[liafa]{M. Habib}
\author[liafa]{V. Limouzy}
\author[liafa]{F. de Montgolfier}
\address[lirmm]{LIRMM, CNRS and University Montpellier II, 161 rue Ada, 34392 Montpellier Cedex 5, France. \texttt{\textup{ buixuan@lirmm.fr}}}
\address[liafa]{LIAFA, CNRS and University Paris Diderot - Paris7, Case 7014, 75205 Paris Cedex 13, France. \texttt{\textup{\{habib,limouzy,fm\}@liafa.jussieu.fr}}}
\begin{abstract}
A new general decomposition theory inspired from modular graph decomposition is presented.
This helps unifying modular decomposition on different structures, including (but not restricted to) graphs.
Moreover, even in the case of graphs, this new notion called \textit{homogeneous modules} not only captures the classical graph modules but also allows to handle $2-$connected components, star-cutsets, and other vertex subsets.

\noindent
The main result is that most of the nice algorithmic tools developed for modular decomposition of graphs still apply efficiently on our generalisation of modules.
Besides, when an essential axiom is satisfied, almost all the important properties can be retrieved.
For this case, an algorithm given by Ehrenfeucht, Gabow, McConnell and Sullivan \cite{EGMcCS94} is generalised and yields a very efficient solution to the associated decomposition problem.
\end{abstract}
\end{frontmatter}
\section{Introduction}\label{sec_intro}
Modular decomposition has arisen in different contexts as a very natural operation on many discrete structures such as graphs, directed graphs, 2-structures, automata, boolean functions, hypergraphs, and matroids.
In graph theory, modular decomposition  plays a central role.
Not only modular graph decomposition yields a framework for the computation of all transitive orientations of a given
 comparability graph~\cite{Gal67,golumbic04,McCS99}, but it  also highly relates to common intervals of 
a set of permutations~\cite{BCMR05,BHP05,M03} and therefore has applications in bioinformatics.
Besides, many graph classes such as cographs, $P_4$-sparse or $P_4$-tidy graphs are characterised by properties of their modules (see e.g.~\cite{BLS99}).
It is also worth noticing that well-known NP-hard problems such as colouring can be solved in polynomial, and often linear, time when the graph is ``sufficiently'' decomposable~\cite{MR84} using some 
application of the \emph{divide and conquer} paradigm.
Finally, the decomposition is useful for graph drawing~\cite{SMC95}, compact encoding (e.g. with cographs~\cite{CLS81} and $P4-$sparse graphs \cite{JO92a}), and precomputing for graph problems including recognition, decision, and combinatorics optimisations (see \cite{MR84} or \cite{BLS99} for a survey).
A central point of this theory relies on the decomposition theorem which presents a tree, so-called \emph{modular decomposition tree}, as compact encoding of the family of modules of a graph.
Then, computing this tree efficiently given the graph has been an important challenge of the past three decades~\cite{BHP05,C97thesis,CHM02,CHM81,CH94,DGMcC01,EGMcCS94,HMP04,HPV99,McCS99,MR84,M03}.
\begin{figure}[t]
~\hfill\includegraphics[width=0.99\textwidth]{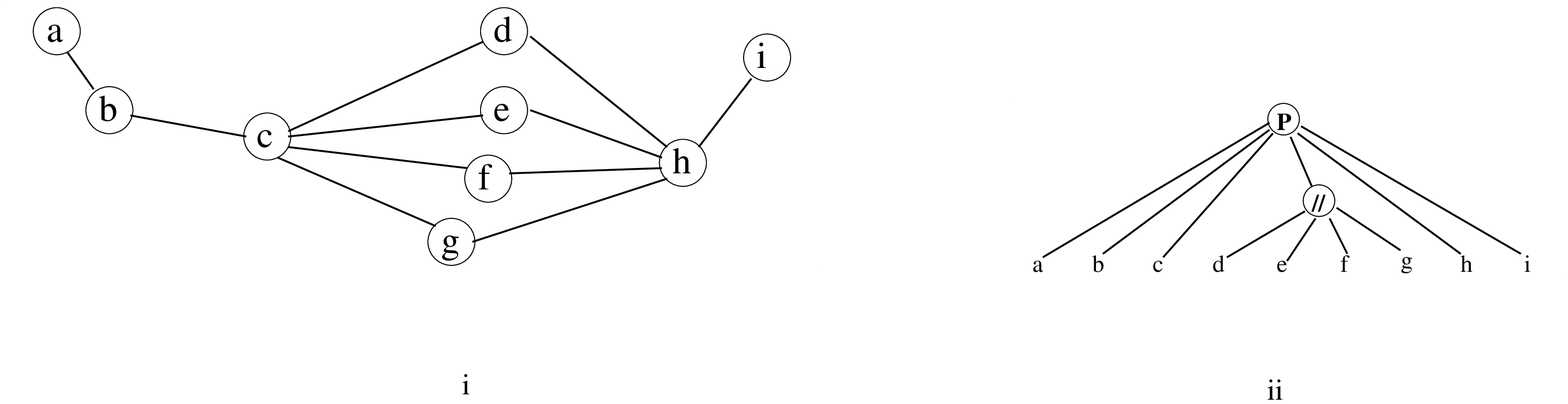}\hfill~
\caption{\label{fig_intro} Illustration of classical modular graph decomposition.
i. In this undirected graph, vertex $c$ is a splitter of $\{a,b\}$ (not linked the same way), whereas vertex $b$ is not a splitter of $\{a,c\}$.
Vertex set $\{d,e,f,g\}$, as well as any of its subsets, is a module of this undirected graph.
ii. Modular decomposition tree of the graph.
}
\end{figure}

On the other hand, several combinatorial algorithms are based on partition refinement techniques~\cite{HMPV00,HPV99,MR917035}.
Many graph algorithms  make intensive use of \emph{vertex splitting}, the action of splitting parts according to the neighbourhood of a vertex.
For instance, all known linear-time modular decomposition algorithms on graphs use this technique~\brochette.
In bioinformatics also, the distinction of a set by an element, so-called \emph{splitter}, seems to play an important role, e.g. in the efficient computation of the set of common intervals of two permutations \cite{BHP05,UY00}.

An abstract notion of \emph{splitter} is studied here and  a formalism based on the concept of homogeneity is proposed.
The resulting structures will be referred to as \emph{homogeneous relations}.
Our aim is a better understanding of existing modular decomposition algorithms by characterising the algebraic properties on which they rely.
As a natural consequence, the new formalism unifies modular decomposition on graphs and on their common generalisations to directed graphs~\cite{McCFM05} and to $2-$structures~\cite{ER90}.
Of course, the theory still applies on structures beyond the previous ones.
Moreover, even in the case of graphs, this new notion called \textit{homogeneous modules} not only captures the classical graph modules but also allows to handle other vertex subsets, e.g. those similar to $2-$connected components, or to star-cutsets.

Our main result is that most of the nice algorithmic tools developed to compute the modular decomposition tree of a graph still apply efficiently in the general theory. 
For graph modules, to design efficient algorithms there actually are three main approaches, distinguishable by the use of properties of:
the set of maximal modules excluding a vertex~\cite{EGMcCS94},
a factoring permutation~\cite{BHP05,CHM02,HMP04,HPV99},
or the visit order of some peculiar graph search such as the so-called  LexBFS  \emph{lexicographic breadth-first search}~\cite{BretscherCHP03,DH89,HM91}.
Because of its specificity due to exotic graph searches, the use of the third approach in the new theory is forfeit.
Still, we extend the two first approaches, and retrieve most of the common efficient computations.

However, as a consequence of their broadness, no obvious decomposition theorem, to our knowledge, is available for arbitrary homogeneous relations, hence no homogeneous modular decomposition tree necessarily is guaranteed.
Indeed, though the homogeneous modules inherit many interesting properties from graph modules, they do not necessarily satisfy the following essential one.
One can \emph{shrink} a whole graph module $M$ into one single vertex $m\in M$: if some vertex of $M$ distinguishes two exterior vertices, then so does every vertex of $M$ and so does $m$.
Let us denote the property by the name of \emph{modular quotient}.
It actually is the basis of many divide-and-conquer paradigms derived from the modular graph decomposition framework, such as the computation of weighted maximal stable or clique set, and graph colouring~\cite{GR97,Moh85}.
This naturally motivates us to study homogeneous relations fulfilling the modular quotient property, hereafter denoted by \emph{good homogeneous relations}.
As expected, almost all important properties of modular graph decomposition, including the decomposition theorem, still hold for the latter relations.
Eventually, we generalise an algorithm given by Ehrenfeucht et al.~\cite{EGMcCS94} to an $O(|X|^2)$ algorithm computing the decomposition tree of a given good homogeneous relation on $X$.

The paper is structured as follows.
First the new combinatorial decomposition theory is detailed  in Sections~\ref{sec_theory} and~\ref{sec_partitive}.
Section~\ref{sectalgo} investigates the general algorithmic framework on arbitrary homogeneous relations.
The subsequent Section~\ref{sec_ghr} is devoted to good homogeneous relations.
Finally, we close the paper with noteworthy outcomes.

\section{Homogeneity, an abstraction of Adjacency}\label{sec_theory}
Throughout this section $X$ is a finite set, and $\mP(X)$ denotes the family of all subsets of $X$. 
A \emph{diverse} triple is $(x,y,z)\subseteq X^3$ with $x\ne y$ and $x\ne z$.
This will be denoted by $(x|yz)$ instead of $(x,y,z)$ since the first element plays a particular role.
Let $H$ be a relation over the diverse triples of $X$.
Given $x\in X$, we define $H_x$ as the binary relation on $X\setminus\{x\}$ such that $H_x(y,z)\Leftrightarrow H(x|yz)$.
\begin{definition}[Homogeneous Relation]\label{def_hr}
$H$ is a \emph{homogeneous relation on $X$} if, for all $x\in X$, $H_x$ is an equivalence relation on $X\setminus\{x\}$
(i.e. it fulfils the symmetry, reflexivity and transitivity properties).
Equivalently, such a relation can be seen as a mapping from each $x\in X$ to a partition of $X\setminus\{x\}$, namely the equivalence classes of $H_x$.
\end{definition}
\begin{definition}[Homogeneous Module]\label{def1}
Let $H$ be a homogeneous relation on $X$.
A subset $M\subseteq X$ is a \emph{homogeneous module} of $H$ if

~\hfill$\forall m,m' \in M, \ \ \forall x\in X\setminus M, \ \ H(x|mm').$\hfill~
\end{definition}

\remark From the definition it is obvious that, given a homogeneous  module $M$, if $\neg H(x|mm')$ for some $m,m'\in M$ then $x\in M$.

If $\neg H(x|mm')$ we say that $x$ \emph{distinguishes} $m$ from $m'$, or $x$ is a \emph{splitter} of $\{m,m'\}$.
A homogeneous module $M$ is \emph{trivial} if $|M|\le 1$ or $M=X$.
The family of homogeneous modules of $H$ is denoted by $\mM_H$, and $\mM$ when no confusion occurs.
$H$ is \emph{modular prime} if $\mM_H$ is reduced to the trivial homogeneous modules.
For convenience, such a relation is also called \emph{prime} when it clearly appears in the context that modules are involved.
Homogeneity and distinction can be applied to graphs.
Indeed, there is a natural homogeneous relation associated to graphs as follow.
\begin{definition}[Standard Homogeneous Relation]\label{def:cano}
The \emph{standard homogeneous relation} $H(G)$ of a directed graph $G=(X,A)$ is defined such that, for all $x,u,v\in X$, $H(G)(x|uv)$ is true if and only if the two following conditions hold:\\
1. either both $u$ and $v$ or none of them are in-neighbours of $x$, and\\
2. either both $u$ and $v$ or none of them are out-neighbours of $x$.
\end{definition}
\begin{figure}[t]
~\hfill\includegraphics[width=0.4\textwidth]{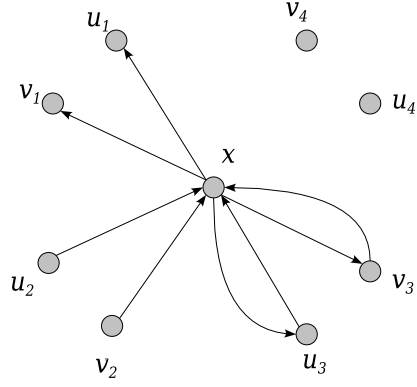}\hfill~
\caption{\label{fig_standard_hr} The standard homogeneous relation $H$ of this directed graph satisfies $H(x|u_iv_i)$ for all $i$, and $\neg H(x|u_iv_j)$ for all $i\neq j$.}
\end{figure}

Roughly, $ H(x|uv)$ tells if $x$ ``sees'' $u$ and $v$ the same way.
Of course the above definition also holds for undirected graphs, tournaments, oriented graphs, and can also be extended to $2-$structures (which roughly are edge-coloured complete directed graphs $G=(X,X^2)$, see e.g.~\cite{ER90} for further information).
It follows straight from definition that
\begin{proposition}\label{propo:cano}
Let $G$ be a graph, resp. tournament, oriented graph, directed graph, $2-$structure.
Homogeneous modules of its standard homogeneous relation $H(G)$ are modules of $G$ in the usual sense \cite{ER90,Gal67,MR84}.
\end{proposition}

Standard homogeneous relations are closely related to the notion of adjacency in graph theory.
Notice that there are other homogeneous relations bound to a graph or to a $2-$structure (e.g. in Section~\ref{sec_outcomes}).
Let us now give some first structural properties of homogeneous relations.
Given $A\subseteq X$ one can define the induced relation $H[A]$ as $H$ restricted to diverse triples of $A^3$.
If $A$ is a homogeneous module we have the following nice property:
\begin{proposition}[Restriction]\label{propo_restriction}
Let $H$ be a homogeneous relation, $M$ a homogeneous module of $H$, and $N\subseteq M$.
Then, $N\in \mM_{H[M]}\ \equi \  N\in \mM_H$.
\end{proposition}
\begin{proof}
That a homogeneous  module of $H$ is a homogeneous  module of $H[M]$ is straight  from definition.
Conversely, if $N\subseteq M$ is not a homogeneous module of $H$, then there is a splitter $s\in X\setminus N$ such that $\exists x,y\in N, \neg{H(s|xy)}$.
However, $s$ cannot belong to $X\setminus M$ since this would imply $s$ is a splitter w.r.t. $H$ of $M$.
Therefore, $s\in M\setminus N$, and is a splitter w.r.t. $H[M]$ of $N$.
Hence, $N$ is not a homogeneous module of $H[M]$.
\end{proof}

\subsection{Lattice Structure}
Let $H$ be an arbitrary homogeneous relation over a finite set $X$.
Let $\mM$ denote the family of its homogeneous modules.
Two sets $A$ and $B$ \emph{overlap} if $A\cap B$, $A\setminus B$ and $B\setminus A$ all are non-empty.
It is denoted by $A\chev B$.
\begin{proposition}\label{prop1}
$\forall A, B\in \mM$, if $A\chev B$, then $(A\cap B)\in \mM$ and $(A\cup B)\in \mM$.
\end{proposition}
\begin{proof}
the fact that $A\cap B$ is a homogeneous module is obvious.
We use the transitivity of $H_x$ for all $x\notin A\cup B$ to prove $(A\cup B)\in \mM$.
\end{proof}
\begin{proposition}
If $\mM$ denotes the family of homogeneous modules of a homogeneous relation, and $\mM'=\mM\cup\{\emptyset\}$, then $(\mM',\subseteq)$ is a lattice.
\end{proposition}
\begin{proof}
Since $\emptyset\in\mM'$, and thanks to Proposition~\ref{prop1}, the intersection of two members $A$ and $B$ belonging to $\mM'$ belongs to $\mM'$.
It is the infimum of $A$ and $B$, since any member of $\mM'$ that is a subset of both $A$ and $B$ is a subset of $A\cap B$.
Let $\mN$ be the family of all members of $\mM'$ containing both $A$ and $B$.
It is non-empty for $X$ is a member.
Since $\mM'$ is closed under intersection, $\mN$ admits a unique smallest member (w.r.t. inclusion), which is the intersection of all its members, and is the supremum of $A$ and $B$.  
\end{proof}

This lattice is a sublattice of the boolean lattice (hypercube) on $X$.
Moreover, if we consider $A \in \mM$ such that $|A| \geq 1$, and $\mM(A)= \{M \in \mM_H~and~M \supseteq A\}$, then $(\mM(A),\subseteq)$ is a distributive lattice.

\subsection{Homogeneous Modules as Roots of a Submodular Function}\label{sec_submod}
Submodular functions are combinatorial objects with powerful potential (see e.g.~\cite{F91}).
Theorem~\ref{theo_submod} below enables the application of this theory to homogeneous relations: the homogenenous modules of any such relation coincide with the roots of a function which satisfies the submodular inequality on intersecting subsets.
\begin{definition}
A set function $\mu:~\mP(X)\rightarrow\mathbb{R}$ is submodular if, for all sets $A,B\in\mP(X)$, $\mu(A)+\mu(B)\geq\mu(A\cup B)+\mu(A\cap B)$ (see e.g. \cite{F91}).
\end{definition}
\begin{theorem}\label{theo_submod}
Let $H$ be a homogeneous relation on $X$.
Let $s(A)$ be the function counting the number of splitters of a non-empty subset $A\subseteq X$.
Then, $s$ follows the submodular inequality on intersecting subsets:
$$s(A)+s(B)\geq s(A\cup B)+s(A\cap B)\textrm{ for all }A\cap B\neq\emptyset.$$
\end{theorem}
\begin{proof}
If $A\subseteq B$ or $B\subseteq A$, the inequality is trivial.
If $A\chev B$ then $A\neq\emptyset$ and $B\neq\emptyset$.
Let $\euss_A$ denote the set of splitters of $A$.
If $\{X_1,\dots,X_k\}$ is a partition of $X$, we note $X=\{X_1,\dots,X_k\}$.
Obviously, ${\EuScript S}_{A\cap B}=\left\{{\EuScript S}_{A\cap B}\setminus B,~{\EuScript S}_{A\cap B}\cap B\right\}$.
As ${\EuScript S}_A\cap A=\emptyset$, the partition ${\EuScript S}_{A\cup B}=\{{\EuScript S}_{A\cup B}\setminus {\EuScript S}_A,~{\EuScript S}_{A\cup B}\cap{\EuScript S}_A\}$ can be reduced to
${\EuScript S}_{A\cup B}=\{{\EuScript S}_{A\cup B}\setminus {\EuScript S}_A,~{\EuScript S}_A\setminus (A\cup B)\}$.
Similarly,
${\EuScript S}_B=\left\{{\EuScript S}_B\setminus {\EuScript S}_{A\cap B},~{\EuScript S}_{A\cap B}\setminus B\right\}.$
Finally,
${\EuScript S}_A=\{{\EuScript S}_A\setminus B,~({\EuScript S}_A\cap B)\setminus {\EuScript S}_{A\cap B},~({\EuScript S}_A\cap B)\cap{\EuScript S}_{A\cap B}\}$
can be reduced to
${\EuScript S}_A=\{{\EuScript S}_A\setminus(A\cup B),~({\EuScript S}_A\cap B)\setminus {\EuScript S}_{A\cap B},~{\EuScript S}_{A\cap B}\cap B\}$. Hence,

$|{\EuScript S}_A|+|{\EuScript S}_B|-|{\EuScript S}_{A\cup B}|-|{\EuScript S}_{A\cap B}|=|({\EuScript S}_A\cap B)\setminus {\EuScript S}_{A\cap B}|+|{\EuScript S}_B\setminus {\EuScript S}_{A\cap B}|-|{\EuScript S}_{A\cup B}\setminus {\EuScript S}_A|.$

To achieve proving the theorem, we prove that ${\EuScript S}_{A\cup B}\setminus {\EuScript S}_A\subseteq{\EuScript S}_B\setminus {\EuScript S}_{A\cap B}$.
Indeed, let $s\in{\EuScript S}_{A\cup B}\setminus{\EuScript S}_A$.
Then, $s\notin A\cup B$ and $H(s|xy)$ for all $x,y\in A$.
Now, suppose that $s\notin\euss_B$.
Since $s$ does not belong to $B$, we deduce $H(s|xy)$ for all $x,y\in B$.
Furthermore, as $A$ and $B$ overlap and thanks to the transitivity of $H$, we deduce $H(s|xy)$ for all $x,y\in A\cup B$ and $s\notin A\cup B$, which is by definition $s\notin\euss_{A\cup B}$. Contradiction.
Finally, supposing $s\in\euss_{A\cap B}$ would imply $s\in\euss_A$.
\end{proof}

In~\cite{UY00} a (restricted) version of this theorem is proved, and this submodularity property is used to propose a very nice algorithm which computes the set of common intervals of a set of permutations.
This approach was generalised for modules of standard homogeneous relations of undirected graphs in~\cite{BHP05}.
It would be interesting to consider this idea on arbitrary homogeneous relations.

\subsection{Strong Homogeneous Modules and Primality}\label{sec_gdt}
In an arbitrary family $\mF$ of subsets of $X$, a member $A\in\mF$ is \emph{strong} if it does not overlap any other member $B\in\mF$.
Those which are not strong are \emph{weak}.
If they belong to the family, $X$ and the singletons $\{x\}$ $(x\in X)$ form the \emph{trivial} strong members of $\mF$.
Otherwise we extend $\mF$ with the trivial strong members.

The set inclusion orders the strong members of $\mF$ into a tree, hereafter denoted by the \emph{generalised decomposition tree of $\mF$}.
This could be seen as a quick proof that, in $\mF$, there are at most $2|X|-1$ strong members, and at most $|X|-2$ non-trivial ones, since the tree has $|X|$ leaves and no degree $2$ internal nodes, except for possibly the root.
When $\mF$ is \emph{weakly partitive} (see definition in Section~\ref{sec_partitive}), this tree plays an important role since it is an exact coding in $O(|X|)$ space of the possibly $2^{|X|}$ members of the family.
It is then called the \emph{decomposition tree of $\mF$}.

The \emph{parent} of a (possibly weak) member $M\in\mF$ is the smallest strong member $M_P$ properly containing $M$, and $M$ is said to be a \emph{child} of $M_P$.
For instance, if $M$ is strong then $M_P$ is its parent in the generalised decomposition tree.
A strong member is \emph{prime} if all its children are strong, and \emph{brittle} otherwise.
\begin{figure}[t]
~\hfill\includegraphics[width=0.5\textwidth]{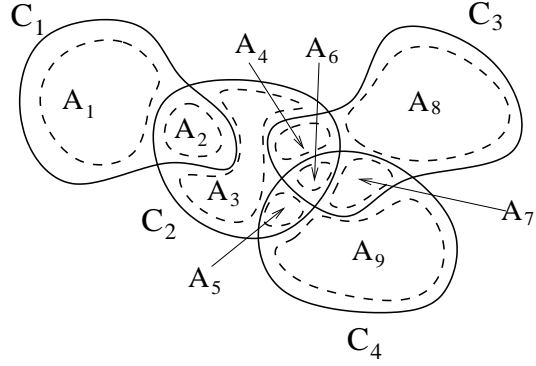}\hfill~
\caption{\label{fig_atoms} The atoms $A_1,\dots,A_9$ of the overlap class $\mC=\{C_1,C_2,C_3,C_4\}$.}
\end{figure}

An \emph{overlap class of $\mF$} is an equivalence class of the transitive closure of the overlap relation $\chev$ on $\mF$.
Such a class is \emph{trivial} if it contains only one member $A\in\mF$.
Then $A$ is by definition a strong member of $\mF$.
The \emph{support} of an overlap class $\mC=\{C_1,\dots,C_k\}$ is defined as $S(\mC)=C_1\cup\dots\cup C_k$.
An \emph{atom} of the overlap class $\mC$ is a maximal subset of $S(\mC)$ that does not overlap any $C_i$ $(1\leq i\leq k)$ (an illustration is given in Fig.~\ref{fig_atoms}).
Notice that the atoms form a partition of $S(\mC)$.
Besides, an atom of an overlap class belongs to the class if and only if this class is trivial.
Furthermore, the support, resp. an atom, of an overlap class belongs to the family $\mF$ if and only if it is a strong member of $\mF$.
Of course, a support, resp. an atom, does not necessarily belong to $\mF$.
However, in a \emph{weakly partitive} family (see Section~\ref{sec_partitive}), all atoms and supports of overlap classes will belong by definition of partitivity to $\mF$, hence are strong members of the family.
It is an elementary result of finite set theory that
\begin{proposition}\label{propo_over}
The following holds for any family $\mF$ of subsets of a finite set $X$ satisfying the closure under union of overlapping members.\\
1. $A\subseteq X$ is a prime strong member of $\mF$ if and only if $\{A\}$ is a trivial overlap class of $\mF$.\\
2. $A\subseteq X$ is a brittle strong member of $\mF$ if and only if it is the support of some non-trivial overlap class $\mC_A$ of $\mF$.
In this case, weak children of $A$ coincide with members of $\mC_A$.
\end{proposition}

Of course we apply all these notions to the family of homogeneous modules of a homogeneous relation $H$.
Let $Z(x,y)$ be the largest homogeneous module of $H$ containing $x$ but not $y$. 
$Z(x,y)$ is well defined since it is the union of all homogeneous modules containing $x$ but not $y$, which is a homogeneous  module thanks to Proposition~\ref{prop1}.
Moreover, $Z(x,y)$ is not empty because $\{x\}$ is a member.
Let $\mZ(H)$ be the family
$$\mZ(H)=\{Z(x,y)~|~~x,y\in X~\wedge~x\ne y\}.$$

Notice that $\mZ(H)$ is not necessarily closed under union of overlapping members.
An example of such $\mZ(H)$ is as follows.
If $X=\{a,b,c\}$, $H(a|bc)$, $H(b|ac)$, and $H(c|ab)$, then $\{a,b\}\in\mZ(H)$, $\{a,c\}\in\mZ(H)$, however $X\notin\mZ(H)$.
\begin{theorem}\label{thfab}
All support and atoms of $\mZ(H)$ that are homogeneous modules of $H$ are strong homogeneous modules.
A non-trivial strong homogenous module of $H$ is either the support or an atom of some overlap class of $\mZ(H)$. 
\end{theorem}
\begin{proof}
Let us prove the first claim of the theorem.
\begin{enumerate}
\item The support of  an overlap class of $\mZ(H)$ is a homogeneous module, since the
family of homogenous modules is closed under the union of overlapping members
(Proposition~\ref{prop1}).  If the support $S$ of a given overlap
class $\mC$ is overlapped by another homogenous module, then it is overlapped by a homogenous module $A\notin
\mZ(H)$. Let $x$ be an element of $A\setminus S$ and $y$ an element of
$S\setminus A$. $Z(x,y)$ contains $A$ but not $\{y\}$ and thus overlaps $S$, so it must overlap at least one member 
of $\mC$ and thus $Z(x,y)\in C$, a contradiction since $x\notin S$. So the support of an overlap class is a strong homogenous module.
\item
Let $A$ be an atom of a given overlap class $\mC$ of $\mZ(H)$. If $A$ is included in
at least two members of $\mC$, then $A$ is exactly the intersection of
all members of $\mC$ which include $A$.  Since the family of homogenous modules is
closed under intersection of overlapping members
(Proposition~\ref{prop1}), $A$ is a homogenous module. Notice that if $A$ is
included in only one member of $\mC$, it may fail to be a homogenous module.  Let
us suppose that $A$ is a homogenous module, and that it is overlapped by another
homogenous module. Then it is overlapped by a homogenous module $B\notin \mZ(H)$. Let $x$ be an element
of $B\setminus A$ and $y$ an element of $A\setminus B$. $Z(x,y)$
contains $B$ but not $\{y\}$ and thus overlaps $A$, so it overlaps all elements of $\mC$
which include $A$ and thus $Z(x,y)\in\mC$, a
contradiction since no atom may be overlapped by a member of the overlap class. So the atoms of an overlap class which are homogenous modules are strong.
\end{enumerate}

Now, let us prove that if $M$ is a non-trivial strong homogenous module then it is the support or an atom of some overlap class.
We shall distinguish three cases.
Let $M_P$ be the strong parent of $M$ (which exists since $M\ne X$).
\begin{enumerate}
\item $M$ is prime and $M_P$ is prime.
Then for all $x\in M$ and all $y\in M_P\setminus M$, $M=Z(x,y)$.
As $M$ is a strong homogenous module, it alone forms a trivial overlap class of $\mZ(H)$ and is equal to its support and to its unique atom.
\item $M$ is prime and $M_P$ is brittle.
Then for all $x\in M$ and all $y\in M_P\setminus M$, $M$ is included in $Z(x,y)$.
Notice that these $Z(x,y)$ belong all to a same overlap class $\mC$ of $\mZ(H)$.
Since $M$ is a strong  homogenous module of $H$, $M\subseteq S(\mC)$ cannot overlap any member of $\mC$.
Moreover, for all $M\subsetneq N\subseteq S(\mC)$, $N$ would overlap $Z(x,y)$ with $x\in M$ and $y\in N\setminus M$.
Hence, $M$ is by definition an atom of $\mC$.
\item $M$ is brittle. It is easy to notice that  $M$ has $k\ge 3$ strong children $M_1,\dots,M_k$.
  Let
us pick an element $x_i$ in each $M_i$.  Then for all $i$ and $j$ we
consider $Z(x_i,x_j)$. Not all of them are strong homogeneous modules (otherwise, $M$ would be prime). 
Let us consider the \emph{overlap graph} of
these homogeneous modules (the vertices are the homogeneous modules, and there is an edge
between overlapping homogeneous modules). Each connected component of this graph
is an overlap class.  According to the first sentence of the theorem,
the support of each overlap class is a strong homogeneous module. If there are two
overlap classes, the support of at least one is a strong homogeneous module that is
strictly between $M$ and its sons $M_i$ in the inclusion tree, since the overlap graph has at least one edge, a
contradiction. So there must be only one overlap class, whose support
is exactly $M$.
\end{enumerate}
\end{proof}

For an arbitrary homogeneous relation, Theorem~\ref{thfab} gives the basis for an $O(|X|^3)$ time enumeration of all strong homogeneous modules, which is depicted in Section~\ref{sectfort}.

\subsection{Particular Homogeneous Relations}\label{sec_special_hr}
We now survey some classes of homogeneous relations defined by added axioms, which, in practice, frequently occurs.
For instance, the class of standard homogeneous relations (see Definition~\ref{def:cano}) has very specific properties, leading to efficient decomposition algorithms (see Section~\ref{sec_ghr}).
\begin{definition}
A homogeneous relation $H$ is said to be
\begin{itemize}
\item {\bf \emph{weakly graphic}} if $H(y|xz)\et H(z|xy)\impl H(x|yz)$ for all $x,y,z\in X$;
\item{\bf \emph{weakly digraphic}} if $H(s|xy)\wedge H(t|xy)\wedge H(y|sx)\wedge H(y|tx) \Rightarrow  H(x|st)$ for all $x,y,s,t \in X$;
\item {\bf \emph{modular quotient}} if $H(x|st)\equi H(y|st)$ for all homogeneous modules $M$ of $H$, for all $x,y\in M$, and $s,t\notin M$.
\end{itemize}
\end{definition}
\begin{proposition}\label{propo_classification}
A weakly graphic homogeneous relation is weakly digraphic.
There are weakly graphic homogeneous relations that are not modular quotient.
There are modular quotient homogeneous relations that are not weakly digraphic, hence not weakly graphic.
\end{proposition}
\begin{proof}
If $H$ is weakly graphic, $H(s|xy)$ and $H(y|sx)$ imply $H(x|sy)$.
Likewise, $H(t|xy)$ and $H(y|tx)$ imply $H(x|ty)$.
Then, $H(x|st)$ by transitivity of $H_x$.
Hence, $H$ is weakly digraphic.
Besides, let $K$ be defined over $X_K=\{x,y,s,t\}$ as
$K_x=\{\{y\},\{s\},\{t\}\}$,
$K_y=\{\{x\},\{s,t\}\}$,
$K_s=\{\{x,y\},\{t\}\}$,
and $K_t=\{\{x,y\},\{s\}\}$.
Then, $K$ is weakly graphic (exhaustive checking on all triplets) but not modular quotient ($\neg{H(x|st)}$ and $H(y|st)$ for the homogeneous module $\{x,y\}$).
Finally, let $L$ be defined over $X_L=\{x,y,s,t,z\}$ as
$K_x=\{\{s\},\{t,y,z\}\}$,
$K_y=\{\{s,t,x\},\{z\}\}$,
$K_s=\{\{x,y\},\{t,z\}\}$,
$K_t=\{\{x,y\},\{s,z\}\}$.
and $K_z=\{\{x\},\{s,t,y\}\}$.
Then, $L$ vacuously is modular quotient as having no homogeneous module, but not weakly digraphic ($x,y,s,t$ form a counterexample).
\end{proof}

The modular quotient property plays an important role in modular decomposition algorithmics.
Indeed, if $H$ is modular quotient, elements in a homogeneous module $M$ of $H$ uniformly perceive a set $A$ not intersecting $M$: if one element of $M$ distinguishes $A$ then so do all.
This, combined with the definition of a homogeneous module, allows to shrink $M$ into a single element, the quotient by $M$, or to pick a \emph{representative element} from the homogeneous module.
Recursion can therefore be used when dealing with homogeneous modules.
The modular quotient and restriction (Proposition~\ref{propo_restriction}) properties were first used in modular decomposition of graphs and are useful for algorithmics~\cite{MR84}.
In this paper, these relations will be qualified as \emph{good homogeneous relations}, and Section~\ref{sec_ghr} is devoted to their study.

Let the \emph{congruence w.r.t. $H$} of an element $x\in X$ stand for the number of equivalence classes of the relation $H_x$.
Then, the \emph{local congruence} of $H$ is the maximum congruence of all elements of $X$.
Homogeneous relations of congruence $2$ plays a special role in graph theory as they include the class of standard homogeneous relations of undirected graphs and tournaments (see next section).
Furthermore, those relations satisfy the following nice property.
\begin{proposition}
Any weakly graphic homogeneous relation $H$ of local congruence $2$ is modular quotient.
\end{proposition}
\begin{proof}
Suppose $H$ weakly graphic and not modular quotient.
Then, there exist $x,y,s,t$ pairwise distinct elements such that $\{x,y\}$ is a homogeneous module, $H(x|st)$, and $\neg H(y|st)$.
Let us prove that we have both $\neg H(y|xs)$ and $\neg H(y|xt)$.
Indeed, suppose w.l.o.g. that $H(y|xs)$.
Then, the transitivity of $H_y$ implies $\neg H(y|xt)$ (for we already have $\neg H(y|st)$).
Besides, since $\{x,y\}$ is a homogeneous module, $H(s|xy)$.
The weakly graphic property implies $H(x|sy)$, and the transitivity of $H_x$ yields $H(x|ty)$.
But then we would have $H(x|ty)$, $H(t|xy)$ ($\{x,y\}$ homogeneous module), and $\neg H(y|xt)$, which is a contradiction with being weakly graphic.
Hence, $\neg H(y|st)$, $\neg H(y|xs)$, $\neg H(y|xt)$, and the congruence of $y$ is at least $3$.
\end{proof}

\subsection{Standard Homogeneous Relations}\label{sec_standard}
Given a (directed) graph, and more generally a $2-$structure, the associated standard homogeneous relation is defined in Definition~\ref{def:cano}.
Such relations are peculiar and satisfy the following fundamental property.
\begin{proposition}\label{propo_stand_modquot}
The standard homogeneous relation of a $2-$structure is modular quotient.
In particular, this result holds for graphs, tournaments, oriented graphs, and directed graphs.
\end{proposition}

Proposition~\ref{propo_stand_modquot} has important algorithmic implications that will be detailed in Section~\ref{sec_ghr}.
Now, the name of \emph{weakly graphic} and \emph{weakly digraphic} homogeneous relations used in the previous section is motivated by Proposition~\ref{propo_mieux} below.
A \emph{symmetric} $2-$structure refers to an edge-coloured clique (the clique is seen as an undirected graph, see e.g.~\cite{ER90} for further information).
\begin{proposition}\label{propo_mieux}
The standard homogeneous relation of a directed graph, resp. a $2-$structure, is weakly digraphic.
The standard homogeneous relation of an undirected graph, resp. a symmetric $2-$structure, is weakly graphic.
\end{proposition}

We now investigate a converse question: given a homogeneous relation $H$ over a finite set $X$, does there exist an undirected graph, or a tournament, admitting $H$ as standard homogeneous relation?
$H$ is defined as a \textbf{\emph{graphic}} homogeneous relation if its local congruence is at most $2$ and if $H[\{a,b,c\}]$ has exactly $0$ or $2$ elements of congruence $2$ for every triple $\{a,b,c\}$.
$H$ is \textbf{\emph{tournamental}} if its local congruence is at most $2$ and if $H[\{a,b,c\}]$ has exactly $1$ or $3$ elements of congruence $2$ for every triple $\{a,b,c\}$.
\begin{theorem}\label{theo_charact_stand}
$H$ is the standard homogeneous relation of an undirected graph if and only if it is graphic.
$H$ is the standard homogeneous relation of a tournament if and only if it is tournamental.
\end{theorem}
\begin{proof}
It is straightforward to check that the standard homogeneous relation of any graph, resp. tournament, is graphic, resp. tournamental.
The converse for graphs can be proved as follows.
Let $H$ be a graphic homogeneous relation over a finite set $X$, and $x\in X$.
Let $C_x$ be one of the possibly two equivalence classes of $H_{x}$ (there always is at least one such class).
We define the matrix $M$ as: $M(x,y)=1$ if $y\in C_x$ and $M(x,y)=0$ otherwise; for all $x'\neq x$, $M(x',y)=1$ if $y\in C_{x'}$ and $M(x,y)=0$ otherwise, where $C_{x'}$ is the equivalence class of $H_{x'}$ containing $x$.
Suppose $M$ not symmetric.
Then, there exists $y\neq z$ both distinct to $x$ such that $M(y,z)=1$ and $M(z,y)=0$.
But then $H[\{x,y,z\}]$ would have exactly $1$ or $3$ elements of congruence $2$.
Therefore, $M$ is a $\{0,1\}$ symmetric matrix and can be seen as the adjacency matrix of some undirected graph $G$.
It is then straightforward to verify that $H$ is the standard homogeneous relation of $G$.
The proof for tournaments is similar.
We use the characterisation that the adjacency matrix of a tournament is a $\{-1,1\}$ anti-symmetric matrix since there are no non-edges and no double arcs.
\end{proof}
\begin{corollary}
It can be tested in $O(|X|^3)$ time if a homogeneous relation $H$ admits a graph $G$ or a tournament $T$ such that $H(G)=H$ or $H(T)=H$.
\end{corollary}
\begin{proof}
First check if all element $x$ has congruence at most $2$.
Then check for all triples the corresponding property of the restricted relation.
\end{proof}

Notice that, if a graphic, resp. tournamental, relation $H$ is given as $|X|$ sets of equivalence classes of $H_x$ (cf Section~\ref{sec_data}), then, the adjacency list representation of the corresponding graph, resp. tournament, can be built in $O(|X|^2)$ time.
Indeed, for graphs one just has to decide which class of the first vertex $v\in X$ represents its neighbourhood.
Then, for any other vertex $u$, the class containing $v$ will be its neighbourhood if $u$ is a neighbour of $v$, and its non-neighbourhood otherwise.
Simply remove the ``non-neighbourhood'' classes (in $O(|X|)$ time each): the other class in each case is the vertex's adjacency list.
A similar construction can be performed for tournaments in the same $O(|X|^2)$ worst case time.

\remark Extending Theorem~\ref{theo_charact_stand} to symmetric $2-$structures is quite straightforward.
It would be interesting to characterise the standard homogeneous relations of directed graphs, and $2-$structures.

\section{Partitivity and Decomposition Theorem}\label{sec_partitive}
A generalisation of modular decomposition, known from \cite{CHM81}, less
general than homogeneous relations but more powerful, is the
\emph{partitive families}.
The \emph{symmetric difference} of two sets $A$ and $B$, denoted by $A\Delta B$, is $(A\setminus B)\cup (B\setminus A)$.
\begin{definition}
A family $\mF\subseteq \mP(X)$ is \emph{weakly partitive} if it contains $X$ and the singletons $\{x\}$ for all  $x\in X$, and is closed under union, intersection and difference of overlapping members, i.e.\\
\indent~\hfill$A\in \mF \et B \in \mF \et A \chev B \impl A\cap B\in \mF \et  A\cup B\in \mF \et   A\setminus B\in \mF.$\hfill~\\
Furthermore a weakly partitive family $\mF$ is \emph{partitive} if it is also closed under symmetric difference of overlapping members:\\
\indent~\hfill$A\in \mF \et B \in \mF \et A \chev B \impl A\Delta B\in \mF.$\hfill~ 
\end{definition}

Let $\mF$ be a weakly partitive family over $X$.
As mentioned before, strong members of $\mF$ can be ordered by inclusion into a tree, so-called generalised decomposition tree (see Section~\ref{sec_gdt}).
In this tree, the child, under the usual parental notion in trees, of an internal node $M$ is by definition a strong member of $\mF$, which is also a strong child of the strong member $M\in\mF$, in the sense of Section~\ref{sec_gdt}.
Besides, a weak child of the node $M$ will refer to the definition of Section~\ref{sec_gdt}.
Let us define three types of strong members of $\mF$, namely three types of nodes of the tree:
\begin{itemize}
\item \emph{prime} nodes which have no weak children,
\item \emph{degenerate} nodes: any union of strong children of the node belongs to $\mF$,
\item\emph{linear} nodes: there is an ordering of the strong children of the node such that a union of them belongs to $\mF$ if and only if they follow consecutively in this ordering.
\end{itemize}
\begin{theorem}\label{theo_dec}\textbf{\cite{CHM81}}
In a partitive family, there are only prime and degenerate nodes.
In a weakly partitive family, there are only prime, degenerate, and linear nodes.
\end{theorem}

The generalised decomposition tree hence is an $O(|X|)$ space coding of the family: it is sufficient to type the nodes into complete, linear or prime, and to order the children of the linear nodes.
It is then called the \emph{decomposition tree} of the family.
From this tree, all weak members of $\mF$ can be outputted by making simple combinations of the strong children of brittle (degenerate or linear) nodes.
Now, the following property states that homogeneous modules of some homogeneous relations are proper generalisations of (weakly) partitive families.
\begin{proposition}\label{prop_wp}
The homogeneous modules of a weakly graphic, resp. weakly digraphic, homogeneous relation $H$ form a partitive, resp. weakly partitive, family.
\end{proposition}
\begin{proof}
Proposition~\ref{prop1} gives the closure by intersection and union of overlapping members.
Let $A\in\mM_H$ and $B\in\mM_H$ be two overlapping homogeneous modules of $H$.
Suppose that there is a splitter $s$ of $A\setminus B$: there are $x,y\in A\setminus B$ such that $\neg H(s|xy)$.
Moreover, $s\in A\cap B$ otherwise it would be a splitter of $A$.
Finally, since $A\chev B$, there exists an element $t\in B\setminus A$.
We have: $H(x|st)$ and $H(y|st)$ and $H(t|sx)$ and $H(t|sy)$ and $H(t|xy)$.
In other words, $H$ is not \emph{weakly digraphic}.
Hence, the family of homogeneous modules of a weakly digraphic homogeneous relation is weakly partitive.
Besides, suppose that $z$ is a splitter of $A\Delta B$.
Then, $z\in A\cap B$ and there exists $x\in A$ and $y\in B$ such that $\neg H(s|xy)$.
Since $H(x|yz)$ and $H(y|xz)$, $H$ is not \emph{weakly graphic}.
Hence, the family of homogeneous modules of a weakly graphic homogeneous relation is partitive.
\end{proof}

As a result, the homogeneous modules of a standard homogeneous relation form a weakly partitive family because such a relation always is weakly digraphic (cf Section~\ref{sec_standard}).
More generally, we will prove in Proposition~\ref{propo_part} that the homogeneous modules of any homogeneous relation that satisfies the modular quotient property (cf Section~\ref{sec_special_hr}), so-called good homogeneous relation, form a weakly partitive family.
Recall that a weakly digraphic homogeneous relation is not necessarily modular quotient (cf Proposition~\ref{propo_classification}).

\section{Algorithms for Arbitrary Homogeneous Relations }\label{sectalgo}
This section considers a given homogeneous relation $H$ over a ground set $X$, and builds tools for computing the generalised modular decomposition tree of $H$.
The best performance to compute this tree in the general case will be given in $O(|X|^3)$ time in Section~\ref{sectfort}.
Notice that the decomposition Theorem~\ref{theo_dec} does not necessarily hold in this section.

\subsection{Data Structures}\label{sec_data}
According to Definition~\ref{def_hr}, a homogeneous relation $H$ can be represented in $O(|X|^2)$ space by an $n\times n$ matrix $A$ of values in $\llbracket 1,n\rrbracket$ as follows.
If $X=\{x_1,\dots,x_n\}$, each equivalence class of the relation $H_{x_i}$ will be assigned a distinct number from $1$ to $n$.
Then, the cell $A_{i,j}$ has value $k$ if and only if $x_j$ belongs to the equivalence class of $H_{x_i}$ having the value $k$.
This representation allows to test in $O(1)$ time whether $H(x_i|x_px_q)$ by checking if $A_{i,p}=A_{i,q}$.
However, retrieving an equivalence class requires an $O(|X|)$ worst case time.

Another alternative is to use the list representation: each element $x\in X$ will be associated to a list of equivalence classes of the relation $H_{x}$.
This list is allowed to ignore one class $C_x$ among the equivalence classes of $H_x$, for instance the largest one.
Thus, the total used space is $O(n+m)$, with $n=|X|$ and $m=\sum_{x\in X}(n-|C_x|)$.
Though this representation allows access in $O(1)$  to an equivalence class of $H_x$ for any  element $x$, testing if $H(x|yz)$ would require $O(n-|C_x|)$.

Notice that for a homogeneous relation, it is straightforward to construct in $O(|X|^2)$ time a list representation given any matrix representation, and conversely.

\textbf{N.B.} Without further specification, all algorithms presented in this paper take matrix representations as input.

\subsection{Smallest Homogeneous Module Containing a Subset}
Let $S$ be a non-empty subset of $X$.
As $\mM_H$ is closed under intersection, there is a unique smallest homogeneous module containing $S$, namely the intersection of all homogeneous modules containing $S$, denoted henceforth by $SM(S)$.
\begin{algorithm2e}[h!]
\caption{Smallest homogeneous module containing $S$} \label{algoSM} 
\dontprintsemicolon 
Let $x$ be an element of $S$, $M:=\{x\}$ and $F:=S\setminus\{x\}$\;
\While{$F$ is not empty}{
  pick an element $y$ in $F$ ; $F :=  F\setminus\{y\}$ ; $M :=  M\cup\{y\}$  \;
  \For{every element $z\notin (M\cup F)$}{
     \lIf{$\neg H(z|xy)$}
         {$F := F \cup \{z\}$\;}
}
}
output $M$ (now equals to $SM(S)$)\; 

\end{algorithm2e}
\begin{theorem}
Algorithm \ref{algoSM} computes $SM(S)$ in $O(|X|.|SM(S)|)= O(|X|^2)$ time. 
\end{theorem}
\begin{proof}
Time complexity is obvious as the \textbf{while} loop runs $|M|-1$
times and the \textbf{for} loop $|X|$ times. The algorithm maintains
the invariant that every splitter of $M$ is in $F$. When $M$ is
replaced by $M\cup\{y\}$, using transitivity of the relation $H_x$, every splitter for $M\cup\{y\}$
either distinguishes $x$ from $y$, or already is in $F$. The algorithm ends
therefore on a homogeneous module that contains $S$, and thus we have
$SM(S)\subseteq M$. If $M\ne SM(S)$ let $s$ be the first element of
$M\setminus SM(S)$ added to $F$ (eventually added to $M$). It
distinguished two elements $x$ and $y$ from $SM(S)$, contradicting
its homogeneity. So $SM(S)=M$.
\end{proof}

\subsection{Maximal Homogeneous Modules Excluding an Element}\label{sectmax}
\begin{proposition}\label{propo_fastoche}
Let $x$ be an element of $X$.
As $\mM_H$ is closed under union of intersecting subsets, there is a unique partition of $X\setminus\{x\}$ into $S_1,\dots,S_k$ such that every $S_i$ is a homogeneous module of $H$ and is maximal w.r.t. inclusion in $\mM_H$.
\end{proposition}
We call $MaxM(x)$ this partition of maximal homogeneous modules excluding $x$, and propose a partition refining algorithm  for its computation.
It is straight from definition that
\begin{lemma}
Every homogeneous module excluding $x$ (especially the maximal ones) is included in some equivalence class of $H_x$.
\end{lemma}

Therefore our algorithm starts with the partition $P=\{H_x^1,\ldots,H_x^k\}$ of equivalence classes of $H_x$.
Then the partition is refined (parts are split) using the following rule.
Let $y$ be an element, called the \emph{pivot}, and $Y$ the part of $P$ containing $y$.

\noindent\textbf{Rule~1} \textsl{
split every part $A$ of $P$, except for $Y$, into $A\cap H_y^1$,\dots,$A\cap H_y^k$}

Notice that a part is broken if and only if its splitters include $y$.
\begin{lemma}\label{lemsplit}
Starting from the partition $P_0=\{H_x^1..H_x^k\}$, the application of  Rule~1 (for any pivot in any order) until no part can be actually split, produces $MaxM(x)$.
\end{lemma}
\begin{proof}
The refining process ends when no pivot can split a part, i.e when every part is a homogeneous module.
Let us suppose one of these homogeneous modules $M$ is not maximal w.r.t. inclusion: it is included in a homogeneous module $M'$, itself included in an equivalence class of $H_x$.
Let us consider the pivot $y$ that first broke $M'$.
It cannot be out of $M'$, as $M'$ is homogeneous module, nor within $M'$, as a pivot does not break its own part.
But $M'$ was broken, contradiction.
\end{proof}

Let us now implement this lemma into an efficient algorithm.
Let $P_i$ be the partition after the $i$th application of Rule~1, $y$ be a given vertex used as pivot, and $Y_i$ the part of $P_i$ containing $y$.
We say that a part $B$ of $P_j$ descends from a part $A$ of $P_i$ if $i<j$ and $A\subset B$.
Clearly, after $y$ is chosen as pivot at step $i$, $y$ does not distinguish any part of $P_i$ excepted $Y_i$.
If $y$ is chosen as pivot after, at step $j>i$, $y$ may only split the parts of $P_{j-1}$ that descend from $Y_i$.
Only these parts have to be examined for implementing Rule~1.
But $Y_j$ itself has not to be examined.

Let us suppose that, for a part $A$, we can split it in $O(|A|)$ time when applying Rule~1 with pivot $y$.
Then the time spent at step $j$ is $O(|Y_i|-|Y_j|)$, the sum of the size of the parts that descend from $Y_i$ save $Y_j$.
The time of all splittings with $y$ as pivot is $O(|X|)$, leading to an $O(|X|^2)$ time complexity.
This is implemented in Algorithm~\ref{algoMaxM}.
\begin{algorithm2e}[ht!]
\caption{Maximal Homogeneous Modules excluding  $x$} \label{algoMaxM} 
\dontprintsemicolon 
\For{ every group $G$}{
\For{ every part $C$ of $G$}{
Compute the set $Z$ of elements in $G$ but not in $C$\;
\For{ every element $y$ of $C$}{
Partition $Z$ according to the equivalence classes of $H_y$\;
Add each partition set to the refining set pool
}}}
Set the group boundaries to the parts boundaries (from $P_{i-1}$ to $P_i$)\;
\For{ each refining set $R$ of the pool}{
Remove $R$ from the pool and then refine  $P_i$ using $R$
}
\end{algorithm2e}

Let us suppose that the parts are implemented as a linked list \cite{HPV99}, and the new parts created after splitting an old one replace it and follow consecutively in the list.
Then for each pivot $y$ two pointers, one on the first part that descends from $Y_i$ and the second to the last part, are enough to tell the parts to be examined.
A simple sweep between the pointers, omitting $Y_j$, gives them. We call all classes descending from a previous one  a \emph{group}.

Now let us show how a part $A$ can be split in $O(|A|)$ time.
It is a classical trick of partition refining \cite{HMPV00,HPV99,MR917035}.
If the equivalence classes of $H_y$ are numbered from $1$ to $k$, then $A$ can be bucket sorted in $O(|A|+k)$ time, then each bucket gives a new part that descends from $A$.
If $|A|<k$, we have to renumber the used equivalence class of $H_y$ from $1$ to $k'\le |A|$ before bucket sorting.
A first sweep on $A$ marks the used equivalence class numbers.
A second sweep unmarks an used number the first time it is seen, and replaces it by the new number (an incremented counter) which is less than $|A|$.
The vector of equivalence class numbers is initialised once in $O(k)$ time.

The last point is the ordering  in which pivots are taken.
Using all elements as pivots, and repeating this $|X|$ times, i.e. $|X|^2$ applications of Rule~1, is enough.
A clever choice is to use $y$ only if $Y_i$ has been split, keeping a queue of ``active'' pivots.
Let us define a measure that will be used later for complexity analysis.
\begin{definition}\label{defQ}
 Let $\mP$ be a partition of $X$.  $Q(\mP)$ be the number of pairs $\{x,y\}$ such that $x$ and $y$ are not in the same part of $\mP$.
\end{definition}

$Q(\mP)$ is between 1 (for the trivial partition $\{X\}$) and $\frac {|X|(|X|-1)}{2}$ (for the trivial partition into singletons).
\begin{theorem}\label{thgraal}
$MaxM(x)$ can be computed in $\Theta(Q(MaxM(x)))=O(|X|^2)$ time. 
\end{theorem}
\begin{proof}
For the correctness proof, one just has to check that the above algorithm implements correctly Lemma~\ref{lemsplit}.
For time complexity issues, notice that, for each pivot $y$, an element $z$ is placed in $Z$ only once. But it is placed in $Z$ only if $y$ and $z$ are not in the same part. 
At each step, refining $Z$ according to the  equivalence classes of $H_y$, and then refining using all sets generated by $y$, takes $O(|Z|)$ time.
Hence the algorithm takes $\Theta(Q(MaxM(x))$ time.
\end{proof}

\subsection{Modular Primality test}
We recall that $H$ is \emph{modular prime} if all its homogeneous modules are trivial (see Section~\ref{sec_theory}).
\begin{theorem}
One can test in $O(|X|^2)$ time if $H$ is modular prime.
\end{theorem}
\begin{proof}
If $|X|\le 2$ the answer is yes.
Otherwise let $x$ and $y$ be two elements of $X$.
In $O(|X|^2)$ time, the algorithm of Section~\ref{sectmax} can output the maximal homogeneous modules excluding $x$.
If one of them is non-trivial then the answer is no.
Otherwise all non-trivial homogeneous modules will contain $x$.
In $O(|X|^2)$ time, the algorithm of Section~\ref{sectmax} can output the maximal homogeneous modules excluding $y$.
If one of them is non-trivial then the answer is no.
Otherwise all non-trivial homogeneous modules will contain $x$ and $y$.
Then, Algorithm~\ref{algoSM} can be used with $S=\{x,y\}$, in $O(|X|^2)$ time.
The answer is yes if and only if $SM(\{x,y\})=X$.
\end{proof}

\subsection{Strong Homogeneous Modules Enumeration}\label{sectfort}
Theorem~\ref{thfab} straightforwardly leads to an algorithm:
\begin{theorem}
The strong homogeneous modules of a homogeneous relation $H$ on $X$ can be enumerated in $O(|X|^3)$ time.
\end{theorem}
\begin{proof}
First compute $MaxM(x)$ for all $x\in X$.
All these sets together form exactly the family $\mZ(H)$ defined in Theorem~\ref{thfab}.
It can be done in $O(|X|^3)$ time using the algorithm of Section~\ref{sectmax} $|X|$ times.
The size of this family (sum of the cardinals of every subset) is $O(|X|^2)$ since they form $|X|$ partitions.
Using Dahlhaus's algorithm \cite{D00} the overlap components can be found in time linear on the size of the family, namely $O(|X|^2)$.
According to Proposition~\ref{propo_over} there are at most $|X|$ non-trivial overlap classes.

For each class it is easy to compute its support, and in $O(|X|^2)$
time easy to compute all its atoms. For instance, consider the vector of
parts of the overlap class containing a given element: the atoms are
the elements with the same vector. Sorting the list of elements of the
supports $O(|X|)$ times, one time per part, gives the elements with
the same vector, thus the atoms.

Then the $O(|X|^2)$ supports and atoms must be sorted by
inclusion order into the inclusion tree of the strong homogeneous modules. It can
be done in $O(|X|^3)$ time using the same sorting technique.

Eventually, ``bad'' atoms -- those that are not strong homogeneous modules -- must be removed from the tree.
According to the first statement of Theorem~\ref{thfab}, the atoms which  are homogeneous modules are strong. We just have to perform 
$O(|X|)$ tests on all nodes of the tree to test which of them are homogeneous modules, which can be done in $O(|X|^2)$ time for each. 
\end{proof}

\subsection{Computation of the Generalised Decomposition Tree given a Factoring Permutation}
The notion of a factoring permutation in the case of graphs~\cite{C97thesis} was introduced to give an alternative for computing the modular decomposition tree of a graph without the precomputing of maximal modules excluding some vertex $x$~\cite{BHP05,CHM02,HMP04,HPV99}.
It can be extended to homogeneous relation as follows.
\begin{definition}[Factoring Permutation]
A \emph{factoring permutation} of a homogeneous relation refers to a depth-first search's visit order of the leaves of the generalised decomposition tree of the relation.
\end{definition}

We here address the problem of, given a homogeneous relation $H$ over a finite set $X$ \emph{and} a factoring permutation $\sigma$, computing the generalised modular decomposition tree of $H$.
Of course the algorithm of Section~\ref{sectfort} answers to this question.
However, this section will depict a more efficient $O(|X|^2)$ solution, which relates to Uno and Yagiura's iterative idea~\cite{BHP05,UY00}.

Actually, the name of factoring permutations is mainly motivated by the following characterisation.
Without loss of generality, we denote the elements of $X$ by $X=\{1,2,\dots,n\}$.
\begin{proposition}
If $\sigma$ is a factoring permutation of a homogeneous relation $H$ over a finite set $X$,
then every strong homogeneous module of $H$ is an interval of $\sigma$, namely it is of the form $\{\sigma(i),\sigma(i+1),\dots,\sigma(j)\}$.
\end{proposition}

Roughly, to enumerate the strong homogeneous modules of $H$, it suffices to find among the intervals of $\sigma$ those that are strong homogeneous modules.
Let $I_{ij}$ denote the $\sigma-$interval $I_{ij}=\{\sigma(i),\sigma(i+1),\dots,\sigma(j)\}$, and $\euss_{ij}$ the splitter set of $I_{ij}$.
\begin{proposition}\label{propo_factp}
$\euss_{ij}=\euss_{(i+1)j}\cup\euss_{i(i+1)}\setminus\{\sigma(i)\}.$
\end{proposition}
\begin{proof}
that $\euss_{(i+1)j}\cup\euss_{i(i+1)}\setminus\{\sigma(i)\}\subseteq\euss_{ij}$ is straight from definition of a splitter.
Conversely, let $x\notin I_{ij}$ be such that $x\notin\euss_{(i+1)j}\cup\euss_{i(i+1)}$.
Then, by the transitivity property of $H_x$, we obtain $H(x|yz)$ for all $y,z\in I_{ij}$, or in other words $x\notin\euss_{ij}$.
Hence, $\euss_{ij}\subseteq\euss_{(i+1)j}\cup\euss_{i(i+1)}$.
We use the fact that $\sigma(i)\notin\euss_{ij}$ to conclude.
\end{proof}

This leads to a naive $O(|X|^3)$ solution to this section's question: for all interval $I_{ij}$, compute $\euss_{i(i+1)}$, then $\euss_{ij}$ using the previously computed $\euss_{(i+1)j}$ and Proposition~\ref{propo_factp}, eventually test if $\euss_{ij}$ is empty.
Let us now improve this idea.
The interval $I_{ij}$ is said to be \emph{right-free} if it does not have a splitter on the right in the order $\sigma$, namely for all $k>j$, $\sigma(k)$ does not belong to $\euss_{ij}$.
Obviously, if $I_{ij}$ is a strong homogeneous module, $I_{ij}$ is right-free.
However, a much more interesting viewpoint is as follows.
If $I_{ij}$ is not right-free, then there will be no $i'\leq i$ such that $I_{i'j}$ is a strong homogeneous module.
Furthermore,
\begin{proposition}
If $j_1<\dots<j_k$ are such that any $I_{ij_q}$ $(1\leq q\leq k)$ is right-free, then $\euss_{ij_1}\subseteq\dots\subseteq\euss_{ij_k}$.
\end{proposition}
\begin{proof}
All splitters of these intervals stand on the left of $\sigma(i)$ in the order $\sigma$.
Hence, a splitter $s$ of $I_{ij_q}$ can not belong to $I_{ij_{q+1}}$, and will belong to $\euss_{ij_{q+1}}$.
\end{proof}

Roughly, if in some iteration step $1\leq i\leq n$, we only store some right-free intervals in a list $RF=(I_{ij_1},\dots,I_{ij_k})$, then all their corresponding splitters can easily be stored by differences in a list $\Delta S=(\Delta_{j_1},\dots,\Delta_{j_k})$, where $\Delta_{j_1}=\euss_{ij_1}$ and $\Delta_{j_q}=\euss_{ij_q}\setminus\euss_{ij_{q-1}}$ $(q\geq 2)$.
Under this convention, an interval $I_{ij_q}$ of the collection is a homogeneous module if and only if all the $q^{th}$ first members of $\Delta S$ are empty: $\Delta_{j_1}=\dots=\Delta_{j_q}=\emptyset$.

From iteration step $i$ to $(i-1)$, the collection of intervals will extend from $RF=(I_{ij_1},\dots,I_{ij_k})$ to $RF=(I_{(i-1)(i-1)},I_{(i-1)j_1},\dots,I_{(i-1)j_k})$, and the list $\Delta S$ will be updated accordingly using Proposition~\ref{propo_factp}.
Also, if for some $j_q$, the extension of $I_{ij_q}$ to $I_{(i-1)j_q}$ introduces a splitter $\sigma(k)$ such that $k>j_q$, then we remove this interval from $RF$ for it no more is right-free and $j_q$ will have no chance to be the right boundary of an unvisited strong homogeneous module.
We come to Algorithm~\ref{algo_quadratic}.
For convenience, each interval $I_{ij_q}$ will be represented by its right boundary: we shall use $RF=(j_1,\dots,j_k)$.
\begin{algorithm2e}[h!]
\caption{Generalised modular decomposition tree computation from a factoring permutation}\label{algo_quadratic} 
\dontprintsemicolon 
\KwIn {a homogeneous relation $H$ over a finite set $X$, and a factoring permutation $\sigma$ of $H$}
\KwOut{the generalised modular decomposition tree $\mT$ of $H$}
$RF\leftarrow()$ and $\Delta S\leftarrow()$ and $M\leftarrow\emptyset$\;
Create a dummy $y=\sigma(n+1)$ such that $H(s|xy)$ for all $s\in X$ and $x=\sigma(n)$\;
\For {$i=n$ downto $1$}{
	$x\leftarrow \sigma(i)$ and $y\leftarrow \sigma(i+1)$\;
	\lIf {$x$ belongs to some member of $\Delta S$}{ remove $x$ from that member\;}
	\For {every $s=\sigma(l)$ with $l<i$ and $\neg{H(s|xy)}$} {
		Add $s$ to the first member of $\Delta S$
	}
	Find $s=\sigma(r)$ such that $\neg{H(s|xy)}$ and $r$ maximum, otherwise $r\leftarrow 0$\;
	\While {the first member $j$ of $RF$ satisfies $j<r$} {
		Remove $j$ from $RF$\;
		Let $Fst$ and $Snd$ be the first and second members of $\Delta S$\;
		$Snd\leftarrow Snd\cup Fst$ and remove $Fst$ from $\Delta S$
	}
	$RF\leftarrow(i,RF)$ and $\Delta S\leftarrow(\emptyset, \Delta S)$\;
	Let $S$, resp. $j$, be the first member of $\Delta S$, resp. $RF$\;
	\While {$S=\emptyset$}{
		$M\leftarrow\{I_{ij}\}\cup M$\;
		Let $S$, resp. $j$, be its next member in $\Delta S$, resp. $RF$\;
	}
}
Remove the weak members of $M$\;
Construct $\mT$, the inclusion order of members of $M$\;
Output $\mT$\;
\end{algorithm2e}

\begin{remark}
Basically, the first step $i=n$ of the main loop still is an initialisation step: at the end of the loop, we always have $RF=(n)$, $\Delta S=(\emptyset)$, and $M=\{n\}$.
The real computation starts at step $i=n-1$.
\end{remark}
\begin{invariant}\label{invar_algoquad}
For all $1\leq i \leq n$, let $RF_i=(j_1,\dots,j_k)$ and $\Delta S_i=(\Delta_1,\dots,\Delta_k)$ be the values of $RF$ and $\Delta S$, at the end of the first loop ``\textbf{for}'' in Algorithm~\ref{algo_quadratic}.
Then,
\begin{itemize}
\item for all member $j$ of $RF_i$, the interval $I_{ij}$ is right-free;
\item for all $1\leq q\leq k$, $\euss_{ij_q}=\Delta_1\cup\dots\cup \Delta_q$.
\end{itemize}
\end{invariant}

Algorithm~\ref{algo_quadratic} correctness directly follows from Invariant~\ref{invar_algoquad}.
As for complexity issues, it is quite straightforward to check that the computing time of all loops is in $O(n^2)$.
After those loops, removing weak members of the list $M$ can be done in linear time on $|M|$ using the lexical member ordering of $M$: $I_{ij}$ is before $I_{i'j'}$ in $M$ if and only if $i\leq i'$ or $(i=i')\wedge (j\leq j')$.
Notice that $|M|$ is less than the number of intervals of $\sigma$, which is in $O(n^2)$.
Likewise, the time spent for ordering by inclusion the remaining members of $M$ is linear on their number using the lexical property.
Whence, the global computing time of Algorithm~\ref{algo_quadratic} is $O(n^2)$.
\begin{theorem}
Given a factoring permutation $\sigma$ of a homogeneous relation $H$ over a finite set $X$, one can compute the generalised modular decomposition tree of $H$ in $O(|X|^2)$ time.
\end{theorem}

Factoring permutations can be get in  $O(|X|^2)$ time in many cases, especially with standard homogeneous relations of
\begin{itemize}
\item inheritance graphs: a linear extension gives a factoring permutation~\cite{DH89};
\item chordal graphs: the cardinality lexicographic breadth first search of the graph yields a factoring permutation~\cite{HM91};
\item tournaments: a very simple partition refining algorithm (greedily choose $x$ and partition the class containing $x$ into $N^-(x),\{x\},N^+(x)$) computes a factoring permutation~\cite{McCFM05};
\item undirected graphs: more sophisticated algorithms run in $O(m\log n)$ time \cite{HPV99} or $O(n+m)$ time \cite{HMP04}.
\end{itemize}

\section{Good Homogeneous Relation Decomposition Algorithm}\label{sec_ghr}
The good homogeneous relations refer to homogeneous relations fulfilling the modular quotient property (cf Section~\ref{sec_special_hr}).
For instance, standard homogeneous relations are good (Proposition~\ref{propo_stand_modquot}).
Their study is motivated by, among others, the following essential property.
\begin{proposition}\label{propo_part}
The homogeneous modules of a good homogeneous relation form a weakly partitive family.
\end{proposition}
\begin{proof}
Proposition~\ref{prop1} gives the closure under intersection and union of overlapping members.
We just have to check that, for two homogeneous modules $A$ and $B$ of $H$, if $A\chev B$ then $A\setminus B$ is a homogeneous module.
Let us suppose $A\setminus B$ has a splitter $s$.
As $A$ is a homogeneous module, $s\in A\cap B$.
Let $x$ and $y$ be two elements of  $A\setminus B$ such that $\neg H(s|xy)$.
As $A\chev B$ there exists $t\in B\setminus A$.
Since $B$ is a homogeneous module, the modular quotient property gives $\neg H(t|xy)$.
But then $A$ no more is a homogeneous module.
\end{proof}

Let $H$ be a good homogeneous relation over a finite set $X$.
We address the problem of computing the modular decomposition tree of $H$, namely the inclusion order of strong homogeneous modules of $H$.
Here again, the algorithm of Section~\ref{sectfort} can be used to give a solution to this question in $O(|X|^3)$ time.
However, this section will give a more efficient $O(|X|^2)$ time solution, which is inspired from Ehrenfeucht et al. works~\cite{EGMcCS94}.
\begin{definition}
A \emph{super-modular-decomposition-tree} (SMDT for short) of a good relation $H$ on $X$ is a tree
\begin{itemize}
\item where the leaf-set is $X$
\item such that each node of the tree is a homogeneous module of $H$
\item such that each strong homogeneous module of $H$ is a node of the tree.
\end{itemize}
\end{definition}
\begin{figure}[b]
~\hfill\includegraphics[width=0.75\textwidth]{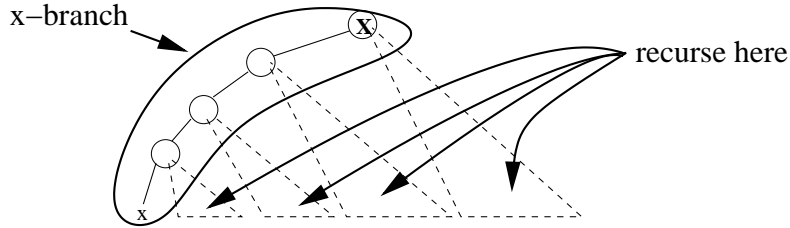}\hfill~
\caption{\label{fig_algo_idea} A recursive approach to compute a super modular decomposition tree.}
\end{figure}

The idea of the algorithm is to compute the left branch (``caterpillar'') of a super modular decomposition tree of $H$, going from the root $X$ to an arbitrary element $x$ (see Fig.~\ref{fig_algo_idea}).
Then, the algorithm recurses to compute the ``legs'' of the caterpillar, and appends them to the caterpillar. Algorithm~\ref{algo_dec} captures this idea.
Eventually, the SMDT is cast into the modular decomposition tree.
\begin{proposition}\label{proppart1}
Algorithm~\ref{algo_dec} computes a super modular decomposition tree
\end{proposition}
\begin{proof}
Obviously all outputted nodes are homogeneous modules. We just have to check that the
tree contains all strong homogeneous modules. This is true indeed, because, for a
strong homogeneous module $M$, the first element $x\in M$ taken for the $x-$branch (see the definition below)
at some recursive step outputs $M\in \mB(x)$. The goodness of the the
relation gives that, when the algorithm is applied recursively on
$H[N]$ and when $N$ is a homogeneous module, the homogeneous module $M$ of $H[N]$ output is
exactly the homogeneous module $M$ of $H$.
\end{proof}
\begin{algorithm2e}[h!]
\caption{Super Modular Decomposition Tree of a Good Homogeneous Relation} \label{algo_dec} 
\dontprintsemicolon 
\KwIn {a good homogeneous relation $H$ over a finite set $X$}
\KwOut{a super modular decomposition tree $\mT$ of $H$}
Let $x$ be an element of $X$\;
Compute $MaxM(x)$, the maximal homogeneous modules not containing $x$\;
Order  $MaxM(x)=M_1..M_k$ such that for each $B_j\in \mB$, $1\le j \le l$,  there exists $f(j)$ such that $B_j = \{x\} \uplus M_1 \uplus M_2  ... \uplus M_{f(j)}$ \;
Initialise $\mT$ to be the $x-$branch\;
\For{every $M_i$ $(1\leq i< k)$}{
	Compute recursively the modular decomposition tree $\mT_i$ of $H[M_i]$\;
	Append $\mT_i$ to the node $B_j$ of $\mT$ such that $j\le i< f(j)$ \;
}
Output $\mT$\;
\end{algorithm2e}

We are now to give a solution to each step of Algorithm~\ref{algo_dec}, and prove their correctness.

\subsection{Strong homogeneous modules containing $x$}
\begin{definition}[$x-$branch]
The $x-$branch of a good homogeneous relation $H$ over $X$ is the set $\mB(x)$
of all strong homogeneous modules containing the element $x\in X$, ordered by inclusion. In other
words, it is the path from the root to leaf $x$ of the modular
decomposition tree of the relation.
\end{definition}

The tool to construct the strong homogeneous modules \emph{containing} $x$ is the construction of the maximal homogeneous modules \emph{excluding} $x$.
Section~\ref{sectmax} defined the set $MaxM(x)=\{M_1,\dots,M_k\}$ of maximal homogeneous modules excluding $x$, which is a partition of $X\setminus\{x\}$ by Proposition~\ref{propo_fastoche}.
Let us examine the relationship between $MaxM(x)$ and $\mB(x)$
\begin{proposition}\label{afirstprop}
The homogeneous modules of $MaxM(x)=\{M_1,\dots,M_k\}$ can be ordered from 1 to $k$ in such a way that 
$$\textrm{for\ each\ } B\in \mB(x),  \textrm{\ there\ exists\ } f \textrm{\ such\ that\ } B = \{x\} \uplus M_1 \uplus M_2  ... \uplus M_f.$$
\end{proposition}
\begin{proof}
For a homogeneous module $B\in  \mB(x)$, the maximal homogeneous modules not containing $B$ form a partition of $X$. Of course each homogeneous module of $MaxM(x)$ is included (or equal to) one of the homogeneous modules of this partition. So a homogeneous module of $MaxM(x)$ can not overlap a homogeneous module  $B\in  \mB(x)$, and the proposition follows.
Indeed, to construct the ordering, just number the homogeneous modules of $\mB(x)$ from $B_0=\{x\}$ to $B_l=X$ using inclusion order. Then number the homogeneous modules  of $MaxM(x)$  included in $B_1$ from 1 to $f(1)$, the homogeneous modules  of $MaxM(x)$  included in $B_2$ but not in $B_1$  from $f(1)+1$ to $f(2)$, and generally the homogeneous modules included in $B_i$ but not in $B_{i-1}$ from $f(i-1)+1$ to $f(i)$.
\end{proof}

A consequence is that, if we order the elements of the $x-$branch from
$B_0=\{x\}$ to $B_l=X$ in increasing inclusion order, then for all
$1\le i< l$ $B_{i+1}\setminus B_i$ is equal to some elements of
$MaxM(x)$ that follow consecutively in the above ordering.
The following fact is obvious.
\begin{proposition}\label{anotherprop}
Let $B_i\in \mB(x)$ be a non-leaf strong homogeneous module containing $x$,
$C_i^1...C_i^{g(i)}$ be its children in the modular decomposition tree
and $j$ such that $C_i^j = B_{i+1}$ is the child containing $x$. If
$B_i$ is \emph{linear} we suppose the children are ordered according
to the linear ordering.
\begin{itemize}
\item If $B_i$ is \emph{prime} then for all $k\ne j$ $C_i^k \in  MaxM(x)$
\item If $B_i$ is \emph{linear} then $\bigcup_{k=1}^{k=j-1} C_i^k \in  MaxM(x)$ and $\bigcup_{k=j+1}^{k=g(i)} C_i^k \in  MaxM(x)$ 
\item If $B_i$ is \emph{complete} then $\bigcup_{k\ne j} C_i^k \in  MaxM(x)$
\end{itemize}
There are no more elements in $MaxM(x)$ than those described above.
\end{proposition}

\subsection{Quotient relation}
Now let us construct a quotient relation. For all $M_i\in MaxM(x)$ let $e_i\in M_i$ be a \emph{representative element} of $M_i$ (an arbitrary element). The \emph{quotient relation} of $H$ by  $MaxM(x)$, denoted $H(x)$, is the relation 
 $$H(x)=H[\{x,e_1,\dots,e_k\}].$$
\begin{proposition}
The quotient relation of $H$ by  $MaxM(x)$ does not depend on the choice of the representative elements for each $M_i$.
\end{proposition}
\begin{proof}
This is because the relation $H$ is good.
\end{proof}
\begin{proposition}\label{prop_Hx}
Every non-trivial homogeneous module of $H(x)$ contains $x$.
\end{proposition}
\begin{proof}
Suppose there is a non-trivial homogeneous module $\cup_{i\in I}\{e_i\}$ of $H(x)$ that excludes $x$.
Then, $|I|\geq2$, and $\cup_{i\in I} M_i$ is a homogeneous module of $H$ that excludes $x$, larger than an element of $MaxM(x)$, a contradiction.
\end{proof}

For $M_i\in MaxM(x)$, let $S(M_i)\in \mB(x)$ be the smallest homogeneous module of $\mB(x)$ containing $M_i$. Using the notations of Proposition~\ref{afirstprop} if $S(M_i)=B_j$ then $i\le j < j(i)$. Proposition~\ref{anotherprop} gives the relationship between $M_i$ and $S(M_i)$ with respect to $S(M_i)$ type (complete, linear or prime).
We say that $e_i\in M_i$ is a \emph{P-element} (resp. \emph{L-element}, \emph{C-element}) if $S(M_i)$ is prime (resp. linear, complete).
Two elements $e_i\in M_i$ and $e_j\in M_j$ are \emph{companion} one of
each other if $S(M_i)=S(M_j)$. Proposition~\ref{anotherprop} tells
that $e_i$ has zero companion if $S(M_i)$ is complete, zero or one if
$S(M_i)$ is linear and at least one if $S(M_i)$ is prime.

\subsection{Forcing graph}
\begin{definition}[Forcing Graph]
Keeping the above notations, the directed \emph{forcing graph} $G(x)=(V,A)$ is defined as
$V=\{e_1,\dots,e_k\}$; and an arc $(e_i,e_j)\in A$ exists if and only if $\neg H(e_j|x,e_i)$.
\end{definition}
\begin{proposition}\label{zz0}
Let $y$ be a vertex of $G(x)$ and $N^*(y)$ the descendants of $y$ in $G(x)$ (including $y$ itself). $N^*(y)\cup\{x\}$ is the smallest homogeneous module of $H(x)$ containing $y$.
\end{proposition}
\begin{proof}
First notice that all nontrivial homogeneous modules of $H(x)$ contain $x$. Then,
if the forcing graph has an edge $(e_i,e_j)$ then any nontrivial
homogeneous module of $H(x)$ containing $e_i$ also contains $e_j$. All descendants
of $y$ in $G(x)$ are thus in any homogeneous module containing $y$ (and $x$).

Now we shall prove that for any set $A$ of vertices of $G(x)$ with no outgoing arc, $A\cup\{x\}$ is a homogeneous module of $H(x)$.
Indeed, for all $u\in A$ and all $v\notin A$ we have $H(v|x,u)$. As $H$ is a transitive relation, then for all $u,u'\in A$ $H(v|u,u')$ and thus $A\cup\{x\}$ is a homogeneous module. So  $N^*(y)\cup\{x\}$ is a homogeneous module of $H(x)$.
\end{proof}

Let $C$ be a  strongly connected component (SCC for short) of $G(x)$. The above proposition gives that all vertices of $C$ are companions. Furthermore we have:
\begin{proposition}\label{zz1}
~~A non-trivial strongly connected components of $G(x)$ is formed by companion P-elements. Conversely a maximal set of  companion P-elements is strongly connected.
\end{proposition}
\begin{proof}
According to Proposition~\ref{anotherprop} there are no companion
$C$-elements and at most two companion $L$-elements. But clearly
there is no arc between them. So a SCC with at least two vertices
contains companion $P$-elements. According to Proposition~\ref{zz0} if
companion $P$-elements were split into two (or more) SCC $C$ and $D$,
then there would be either a homogeneous module of $H(x)$ containing $C$ but not
$D$, or a homogeneous module of $H(x)$ containing $D$ but not $C$. In both case,
the smallest homogeneous module of $H(x)$ containing $C\cup D$ can not be prime.
 \end{proof}
\begin{proposition}\label{zz2}
Two companion $L$-elements are false twins (they share the same neighbourhood and there is no arc between them). Conversely the pairs of false twins are exactly the companion $L$-elements.
\end{proposition}
\begin{proof}
Let $e$ and $e'$ be two companion $L$-elements. The smallest homogeneous module
$M$ of $H(x)$ containing $\{e,e'\}$ is thus a linear homogeneous module $\{e\}\cup
M' \cup\{e'\}$ where $M'$ is the strong homogeneous module son of $M$ in the
modular decomposition tree of $H(x)$. Of course $x\in M'$.  Both
$\{e\}\cup M'$ and $M' \cup\{e'\}$ are homogeneous modules, and the descendants of
$e$ are exactly the descendants of $e'$ and are $M'$, according to
Proposition~\ref{zz0}. Furthermore since $H$ is good, $e$ and $e'$ are
twins.
\end{proof}

According to the Propositions~\ref{zz0}, \ref{zz1} and \ref{zz2} we have:
\begin{proposition}\label{propo_xbranch}
Any linear extension (topological sort) of $G(x)$ will order $MaxM(x)$ into the ordering of Proposition~\ref{afirstprop}.
\end{proposition}

\begin{proposition}\label{proppart2}
The  $x-$branch of $H$ can be computed in $O(Q(MaxM(x)))$ time
\end{proposition}
\begin{proof}
Remind that  $Q(\mP)$ is the number of pairs $\{x,y\}$ whose vertices are not in the same part of a partition $\mP$ (Definition~\ref{defQ}). Let $k$ be the number of parts of $MaxM(x)$. obviously $k^2=O(Q(MaxM(x)))$ and $Q(MaxM(x))=O(|X|^2)$.
The algorithm is
\begin{itemize}
\item
The maximal homogeneous modules $MaxM(x)$ excluding $x$ can be computed in time $O(Q(MaxM(x)))$, according to   Theorem~\ref{thgraal}, using the algorithm of Section~\ref{sectmax}.
\item Then, the vertices of the forcing graph are determined arbitrarily: for all $1\leq i\leq k$ let $e_i\in M_i$.
\item Then, constructing the forcing graph $G(x)$ in $O(k^2)$ time is obvious
\item Then, the topological sort  $G(x)$ in $O(k^2)$ time is also easy.
\item Lastly Proposition~\ref{afirstprop} tells how the ordering of $MaxM(x)$ allow to construct $\mB(x)$. Notice that all companion vertices appear consecutively in the topological sort and are all regrouped to form $B_{i+1}\setminus B_i$.
\end{itemize}
\end{proof}

We thus have:
\begin{theorem}\label{th10}
Algorithm~\ref{algo_dec} computes a super homogeneous modular decomposition tree in $O(|X|^2)$ time.
\end{theorem}
\begin{proof}
This is a direct application of Propositions~\ref{proppart1} and \ref{proppart2}. We just have to show that the sum of all  $O(Q(MaxM(x)))$ time computations is  $O(|X|^2)$. This is true because 
$Q(MaxM(x))$ is the number of pairs $\{x,y\}$ belonging to two elements of $MaxM(x)$. As the algorithm is recursively launched on a homogeneous module of $MaxM(x)$, each pair  $\{x,y\}$ is counted once, in the recursive call of its least common ancestor of the SMDT finally output.
\end{proof}

\subsection{Testing for weak homogeneous modules and typing the nodes}
Now, by constructing recursively $x-$branches, we can build a super homogeneous modular decomposition tree. This tree however is not the modular decomposition tree of $H$ since:
\begin{itemize}
\item Its nodes are not typed complete, linear or prime,
\item It contains all strong homogeneous modules but may also contain weak homogeneous modules.
\end{itemize}
\begin{definition}
Let $N$ be a node of a SMDT of $H$, with sons $S_1,\dots,S_k$, and $e_i\in S_i$ be an arbitrary element. The 
\emph{quotient of $H$ by $N$} is $H[\{e_1,...,e_k\}]$.
\end{definition}
\begin{proposition}\label{proptypq}
The quotient relation of a node $N$ of a SMDT is either
\begin{itemize}
\item type $P$: with no non-trivial homogeneous module,
\item type $L$: the elements can be linearly ordered in such a way that the homogeneous modules of the quotient relations are exactly the intervals of the relation,
\item type $C$: every subset is a homogeneous module.
\end{itemize}
\end{proposition}

If $N$ has $k$ sons, a trivial  $O(k^2)$ time algorithm can test the type and order the elements if needed.
A classical (and easy to prove) result is that
\begin{proposition}\label{propquadratic}
Let $T$ be a tree with $n$ leaves and no node with only one child. Then 
$$\sum_{N\mathrm{\ node \ of \ }T} degree(N)^2 = O(n^2).$$
\end{proposition}
We can therefore perform quadratic-time computations on each node of a SMDT.
A first application of Proposition~\ref{propquadratic} is
\begin{proposition}
Let  $H$ be a good relation on $X$.
It take $O(|X|^2)$ time to compute the quotient relations for all nodes of a SMDT of $H$.
\end{proposition}
\begin{proof}
A bottom-up sweep, keeping one representative per child, builds the representatives. Each quotient relation can then be computed in time linear on its size, i.e. $O(k^2)$.
\end{proof}

A second application of Proposition~\ref{propquadratic} together with
Proposition~\ref{proptypq} gives that the typing of the nodes of a
SMDT  takes $O(|X|^2)$ time. Note that we abusively consider that
weak homogeneous modules have a type.
Then we can look for the weak homogeneous modules, and cast the SMDT into the genuine modular decomposition tree, using:
\begin{proposition}
Let  $H$ be a good relation on $X$,  and  $N$ be a node of a SMDT, and $F$ be its father in the SMDT. $F$ has another son $A$. If $F$ is linear then take $A$ that immediately precedes or follow $N$ in the linear ordering. Take an element $a\in A$. If $N$ is non-trivial it has at least two sons $B$ and $C$. If $N$ is linear then take $B$ its first child and $C$ its last child. Finally take $b\in B$ and $c\in C$. Then

$N$ is a weak homogeneous module if and only if\\
\rightline{$\{a,b\}$ or $\{a,c\}$ is a homogeneous module of $H[\{a,b,c\}]$.}
\end{proposition}
\begin{proof}
If $\{a,b\}$ or $\{a,c\}$ is a homogeneous module $N$ is obviously weak. Conversely if $N$ is weak then it is overlapped by a homogeneous module $N'$. $N$ and $N'$ have thus the same father $F$ in the modular decomposition tree. If $F$ is complete, any union of a son of $F$ included in $N$ plus one not included  $N'$ overlaps $N$. As $b\in N$ and $a\in (F\setminus N)$ we get the result. And if $F$ is linear (any other arc is excluded), then either the first son of $F$ included in $N$ plus the preceding one in the linear order, overlaps $N$, and $\{a,b\}$ is a homogeneous module; or the last son of $F$ included in $N$ plus the following one in the linear order, overlap $N$, and $\{a,c\}$ is a homogeneous module.
\end{proof}

This proposition, together with a third application of Proposition~\ref{propquadratic}, gives that the weak homogeneous modules can be removed from a SMDT in $O(|X|^2)$ time. We finally have
\begin{proposition}
A Super Modular Decomposition Tree of $H$ can be cast into the modular decomposition tree of $H$ in $O(|X|^2)$ time.
\end{proposition}

And, together with Theorem~\ref{th10} we have:
\begin{theorem}
The modular decomposition tree of a good relation $H$ over $X$ can be
built in $O(|X|^2)$ time.
\end{theorem}

\textit{Conjecture:} When the homogeneous relation is given by list representation (see Section~\ref{sec_data}),
the decomposition tree can be built in $O(n+m\log n)$ time, where $n=|X|$ and $m$ the total length of the lists in this representation.

\section{Outcomes}\label{sec_outcomes}
Let us examine in the sequel some of the applications of this
homogeneity theory to modular decomposition of graphs and
2-structures, and to other graph relations.

From Proposition~\ref{propo_mieux} and Section~\ref{sec_partitive}, the modules of an undirected graph and of a symmetric 2-structure form a partitive family, while the modules of a directed graph just form a weakly partitive family.
All know properties of modular decomposition \cite{MR84} can be derived from this result.  An
$O(n^2)$ modular decomposition algorithm can also be derived from
Section~\ref{sec_ghr} algorithm. It runs in optimal time for relations given as matrices (like an adjacency matrix), but it is less efficient than the
existing algorithms for graphs stored using adjacency lists \brochette.

In a graph we can consider different homogeneous relations, for instance the relation \emph{``there exists a path from vertex $x$ to vertex $y$ avoiding the vertex $s$"}, or a more general relation \emph{``there exists a path from $x$ to $y$ avoiding the neighbourhood of $s$"}.
It is easy to see that these two relations fulfil the basic axioms (symmetry, reflexivity and transitivity).
In the first case, the strong hommogeneous modules form a partition (into the 2-vertex-connected components, minus the articulation points).
The second relation is related to decomposition into star cutsets.

Another interesting relation is $D_k(s|xy)$ if $d(s,x)\le k$ and $d(s,y) \le k$, where $d(x,y)$ denotes the distance between $x$ and $y$.
The case $k=1$ corresponds to modular decomposition.
It is worth investigating the general case.

\section{Conclusion}
We hope that  this homogeneity theory will have many other applications and will be useful to decompose automata~\cite{AM03} and boolean functions~\cite{B05}.
Obviously, the algorithmic framework presented here can be optimised in each particular application, as it has been done for modular graph decomposition \brochette.

{\bf\textit{Acknowledgements:}} We are grateful to J. Gustedt for a helpful discussion and his interesting remarks.
We would like to thank the anonymous referees for their suggestions, which greatly improve the paper.
\bibliography{dam06}
\bibliographystyle{plain}

\end{document}